\documentclass[aps,prl,twocolumn,reprint,floatfix, superscriptaddress]{revtex4-1}
\usepackage[utf8]{inputenc}
\usepackage{graphicx,amsmath,amssymb,braket}
\usepackage{color}
\newcommand{\alex}[1]{{\color{red} #1}}

\usepackage[colorlinks=true,citecolor=myblue,linkcolor=myred]{hyperref}

\definecolor{mygrey}{gray}{0.35}
\definecolor{myblue}{rgb}{0.2,0.2,0.8}
\definecolor{myzard}{cmyk}{0,0,0.05,0}
\definecolor{mywhite}{rgb}{1,1,1}
\definecolor{mywhite}{rgb}{1,1,1}
\definecolor{myred}{rgb}{1,0.,0.3}

\begin{document}

\title{Qubit-photon corner states in all dimensions}
\author{Adrian Feiguin}
\affiliation{Department of Physics, Northeastern University, Boston, Massachusetts 02115, US}
\author{Juan José García-Ripoll}
\affiliation{Institute of Fundamental Physics IFF-CSIC, Calle Serrano 113b, 28006 Madrid, Spain}
\author{Alejandro González-Tudela}
\affiliation{Institute of Fundamental Physics IFF-CSIC, Calle Serrano 113b, 28006 Madrid, Spain}
\begin{abstract}
    
    A single quantum emitter coupled to a one-dimensional photon field can perfectly trap a photon when placed close to a mirror. This occurs when the interference between the emitted and reflected light is completely destructive, leading to photon confinement between the emitter and the mirror. In higher dimensions, the spread of the light field in all directions hinders interference and, consequently, photon trapping by a single emitter is considered to be impossible. In this work, we show that is not the case by proving that a single emitter can indeed trap light in any dimension. We provide a constructive recipe based on judiciously coupling an emitter to a photonic crystal-like bath with properly designed open boundary conditions. The directional propagation of the photons in such baths enables perfect destructive interference, forming what we denote as \emph{qubit-photon corner states}. We characterize these states in all dimensions, showing that they are robust under fluctuations of the emitter's properties, and persist also in the ultrastrong coupling regime.

\end{abstract}
\date{\today}

\maketitle

\textit{Introduction.--}
The radiation properties of a quantum emitter can change modifying the photonic environment around it~\cite{purcell56a}. A particularly simple example of this consists in placing an emitter close to a mirror~\cite{cook87a,eschner01a} or to other quantum emitters~\cite{lehmberg70a,lehmberg70b}.  These configurations in free-space already lead to remarkable effects such as lifetime renormalizations or the modification of atomic resonance fluorescence~\cite{dorner02a,beige02a,bushev04a,dubin07a,glaetzle10a}. However, they are ultimately limited by the reduced solid angle of the emitted light that the mirrors or emitters can cover. All these effects are dramatically enhanced when the emitters couple to one-dimensional (1D) photonic fields such as dielectric~\cite{vetsch10a,thompson13a,goban13a,beguin14a,solano17a,lodahl15a} or microwave~\cite{gu17a} waveguides, where, for example, a single atom can perfectly reflect single photons~\cite{shen05a}. These strong interference effects lead to the emergence of bound states in the continuum (BIC)~\cite{hsu16a} with two emitters~\cite{ordonez06a,longhi07a,tanaka06a,zhou08a,gonzaleztudela11a,gonzalezballestero13a,facchi16a,facchi18a}, or a single emitter in front of a mirror~\cite{dong09a,tufarelli13a,tufarelli14a,hoi15a,pichler17a,calajo19a}, in which a single photon becomes localized despite being energetically in the middle of the continuous spectrum. These BICs, which are entangled light-matter states, have experienced a renewed interest because of their possible applications to realize decoherence-free quantum gates~\cite{paulisch16a,kockum18a}, or non-reciprocal photon transport~\cite{fang17a,muller17a,hamann18a}.

Among the different configurations, the one using a single emitter and a mirror~\cite{dong09a,tufarelli13a,tufarelli14a,hoi15a,pichler17a,calajo19a} is especially advantageous since the BICs in that case are insensitive to the energy mismatch between emitters. Exporting this configuration to higher-dimensional systems was generally thought not be possible, since the wavepacket diffraction precludes perfect destructive interference. Here, we show the contrary by proving that indeed a single quantum emitter can perfectly trap light and create a BIC in any dimension. The key idea is to combine the directional emission occurring in 2D and 3D photonic crystal-like baths~\cite{gonzaleztudela17a,gonzaleztudela17b,galve17a,gonzaleztudela18d}, with an adequate design of open boundary conditions. Then, by placing the emitter close to a corner of the photonic bath, its directional emission and the reflection in the boundary generates a high-dimensional BIC that we label as \emph{qubit-photon corner state}. In contrast to the  recently observed topological photon corner states~\cite{benalcazar17a,serra18a,imhof18a,peterson18a,el19a,xie19a,mittal19a,ota19a}, ours can inherit a strong non-linearity from the emitter, and do not require a topologically non-trivial bath. We characterize these states in two and tree dimensions using exact numerical techniques to take into account the retardation effects and the corrections in the ultra-strong coupling regime, where these states acquire a finite lifetime.

\textit{Setup.--} To illustrate the emergence of these states, we use a $d$-dimensional photonic lattice composed by $N^d$ resonators with energy $\omega_a$, that can tunnel to their nearest neighbour at a rate $J$. With these assumptions, the bath energy dispersion ($\omega(\mathbf{k})$) then only depends on the photonic lattice geometry, which determines the number of nearest neighbours resonators ($N_{nn}$). For the emitter, we take a two-level system (qubit) with energy difference $\Delta$, that is locally coupled at a position $\mathbf{x}_0\in \mathbf{R}^{d}$ to the photonic bath. Thus, the full Hamiltonian reads:
\begin{align}
    H &= \frac{\Delta}{2} \sigma^z +\sum_{\mathbf{x}}\omega_a a^\dagger_\mathbf{x} a_\mathbf{x}+ J\sum_{\langle\mathbf{x},\mathbf{y}\rangle} a^\dagger_\mathbf{x} a_\mathbf{y} +g_{x_0} \sigma^x (a_{\mathbf{x}_0} + a_{\mathbf{x}_0}^\dagger)\,. \label{eq:spin-boson}
\end{align}

Notice, that we have kept the full dipole coupling $g_{\mathbf{x}_0}$ between the emitter and the photonic mode. Like this we can study situations in which the coupling is comparatively weak, $g\ll \Delta,\omega_a$ and the rotating-wave approximation (RWA) is justified, replacing $g\sigma^x(a+a^\dagger) \sim g(\sigma^+a + \sigma^-a^\dagger)$, but also the ultra-strong coupling (USC) regime, which occurs when $g/\Delta \geq 10\%$. In this limit,  the physics of the emitter changes substantially when $g \sim W$, where $W=2 N_{nn} J$ is the photon bandwidth.

We are interested in studying the spontaneous emission dynamics, that is, considering that the emitter is initially excited with no photons in the bath, and then study the time dynamics governed by $e^{-i H t}$. In our case, this is a complicated problem because of the high-dimensional nature of the bath and, in the ultra-strong coupling regime, because the number of excitation is not conserved. Thus, before describing the physics, it is worth explaining the two complementary approaches we used to study this problem.

$\bullet$ \textit{Polaron Hamiltonian.--}
Instead of working with~\eqref{eq:spin-boson} directly, we study the unitarily equivalent polaron Hamiltonian~\cite{shi18b}. This transformed model eliminates much of the entanglement between the quantum emitter and the photonic field, leading to renormalized coupling strengths and qubit frequencies. For moderate coupling strengths or finite-bandwidth models, the polaron Hamiltonian has a single excitation limit that describes the spontaneous emission problem that we want to study
\begin{align}
    H_{\mathrm{pol},\mathrm{1D}} &= \frac{\tilde\Delta}{2}\sigma^z (1+ 8 F^\dagger F) + \sum_{\mathbf{x},\mathbf{y}} J_{\mathbf{x}\mathbf{y}} a_\mathbf{x}^\dagger a_\mathbf{y} \\
    \nonumber &+ 2\tilde\Delta (\sigma^+F + \text{H.c.})+\sum_{\mathbf{x}}\omega_a a^\dagger_\mathbf{x} a_\mathbf{x}.
\end{align}

Within that picture, the emitter interacts with a collective coupling operator $F = \sum_\mathbf{x} f_\mathbf{x} a_\mathbf{x}$ with coupling vector $\mathbf{f}=\{f_\mathbf{x}\}_\mathbf{x}$, and has a renormalized frequency $\tilde{\Delta}$. These parameters can be obtained solving self-consistently the following equations
\begin{equation}
    \tilde\Delta = \Delta e^{-2\sum_\mathbf{k} |f_\mathbf{k}|^2},\;
    \mathbf{f} = \frac{1}{J + \tilde{\Delta}} \mathbf{g}.
\end{equation}
The single-excitation polaron adopts a RWA stanza and is therefore amenable to analytical treatment, much like earlier works with regular lattices and point-like interactions\ \cite{gonzaleztudela17a,gonzaleztudela17b}. As a result, the model supports single-photon solutions
\begin{equation}
    \ket{\psi(t)} = \left[\sum_\mathbf{x}\psi(\mathbf{x},t) a_\mathbf{x}^\dagger + c(t) \sigma^+\right]\ket{\downarrow}\otimes\ket{\mathrm{vac}}\,,\label{eq:wavefunction}
\end{equation}
whose photon and qubit components $\psi(\mathbf{x},t)$ and $c(t)$ follow a linear Schr\"odinger equation with $H_{\mathrm{pol},\mathrm{1D}} $, that can be evolved in time using different numerical methods.

$\bullet$ \textit{Chain mapping and DMRG.--}
As an additional benchmark, we also solve the dynamics of the full spin-boson model of Eq.~\ref{eq:spin-boson} using a time-dependent version of the density matrix renormalization group (tDMRG) \cite{white04a,daley04a,feiguin13a,paeckel19a}. To simulate large high dimensional bosonic baths, the non-interacting lattice Hamiltonian is exactly mapped (it is a unitary transformation) onto a 1D chain of free bosons by means of a Lanczos recursion \cite{chin10b,busser13a,allerdt19a}. The consequent dimensional and entanglement reduction makes the new Hamiltonian amenable to DMRG simulations. Remarkably, with only $N$ bosonic modes in the chain we capture well the dynamics of the emitter. As shown in~\cite{paulisch18a}, this mapping can be combined with the polaron transformation to reduce the amount of entanglement, but this was not required for this study.

\begin{figure}[ht]
    \centering
    \includegraphics[width=\linewidth]{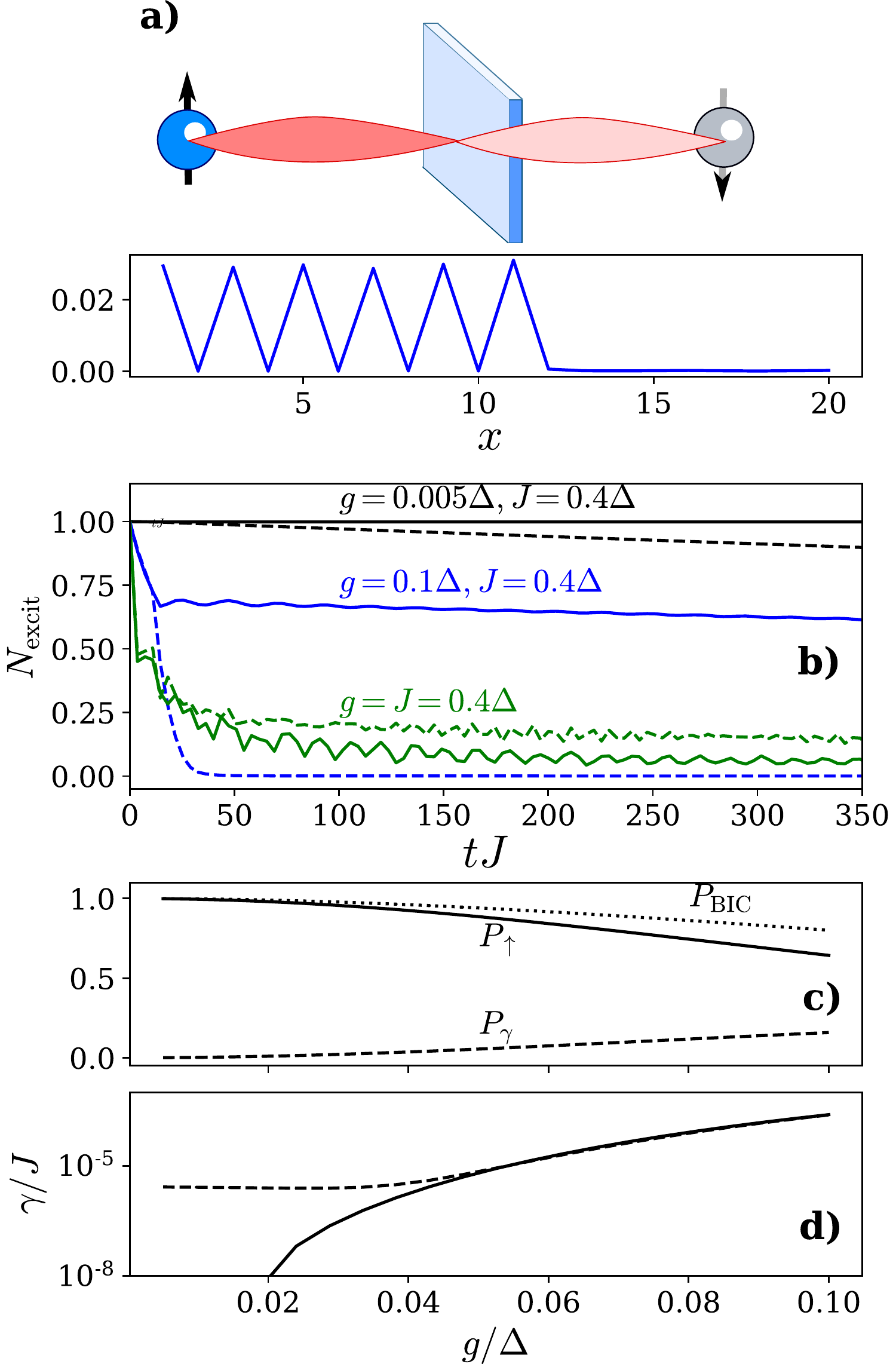}
    \caption{Formation of a 1D BIC by spontaneous emission on a 1D lattice with $400$ sites. (a) Pictorical representation (above) and photon number spatial distribution (below) in the BIC state for $g=0.1\Delta$ and $x=12.$ (b) Total excitation number $N_\text{excit}$\ \eqref{eq:n-excit} in the BIC state as a funciton of time. Solid and dashed lines correspond to $x=12$-th and $x=11$-th site. (c) Qubit and photon component of the bound state, $P_\uparrow$ and $P_\gamma,$ and probability of creating the bound state $P_\text{BIC}\sim N(t_0).$  (d) Estimated decay rate of the corner state extracted from a fit $N_\text{excit}(t)\sim N(t_0)\exp(-\gamma(t-t_0))$ after the initial transient\footnote{Values below $10^{-5}$ are not reliable, due to finite simulation time.}}
    \label{fig:1d-corner}
\end{figure}

\textit{Reminder of 1D BICS.--}
Our first set of simulations recreates the BICs obtained in a one-dimensional lattice with open boundaries and $N=400$ sites, taking the lattice constant as the unit of length. We use an emitter resonant with the middle of the photonic band, $\Delta=\omega_a=2.5J$, although this is not strictly needed. We place a quantum emitter at even ($x=12,$ solid) and odd positions ($x=11,$ dashed), excite the emitter, and abruptly switch on the coupling $g$. When the emitter is placed on an odd site, it decays completely, releasing a propagating photon. However, if the emitter is on an even site, it can, with some probability, trap a photon between the emitter and the end of the lattice, as seen in Fig.~\ref{fig:1d-corner}a.  Such states correspond to the BICs that have been identified before in one-dimensional systems~\cite{ordonez06a,longhi07a,tanaka06a,zhou08a,gonzaleztudela11a,gonzalezballestero13a,facchi16a,facchi18a,dong09a,tufarelli13a,tufarelli14a,hoi15a,pichler17a,calajo19a}, and can be intuitively understood from the interference between the emitted light of the emitter and its afterimage, as schematically depicted in Fig.~\ref{fig:1d-corner}a. Fig.~\ref{fig:1d-corner}b plots the probability of creating the 1D BIC, defined as
\begin{equation}
    N_\text{excit} = \frac{1}{2}(\sigma^z+1) + \sum_{x=1}^{x_0} a^\dagger_x a_x = P_\uparrow + P_\gamma.\label{eq:n-excit}
\end{equation}
which contains both a non-neglibible photonic ($P_\gamma$) and qubit  ($P_\uparrow$) component. Note how the emitter in odd sites decay (dashed lines), but emitters in even sites have some probability to excite the BIC, even in the USC regime. As we increase the coupling strength, the BIC transitions from being mostly an excited atom to an equal superposition of both [cf. Fig.~\ref{fig:1d-corner}c]. In the USC regime, the BIC state has a significant fraction of photon component, but it also acquires a finite lifetime [cf.~Fig.~\ref{fig:1d-corner}d]. This decay can be attributed to the renormalization of the qubit energy when $g\sim W$, which also changes the emitted photon frequencies. Thus, the photons will no longer have the exact wavelength that leads to the perfect interference for the position of the emitter chosen. Finally, note that the results obtained using the single-photon polaron Hamiltonian agree very well with our DMRG simulations.

\begin{figure}[tb]
    \centering
    \includegraphics[width=\linewidth]{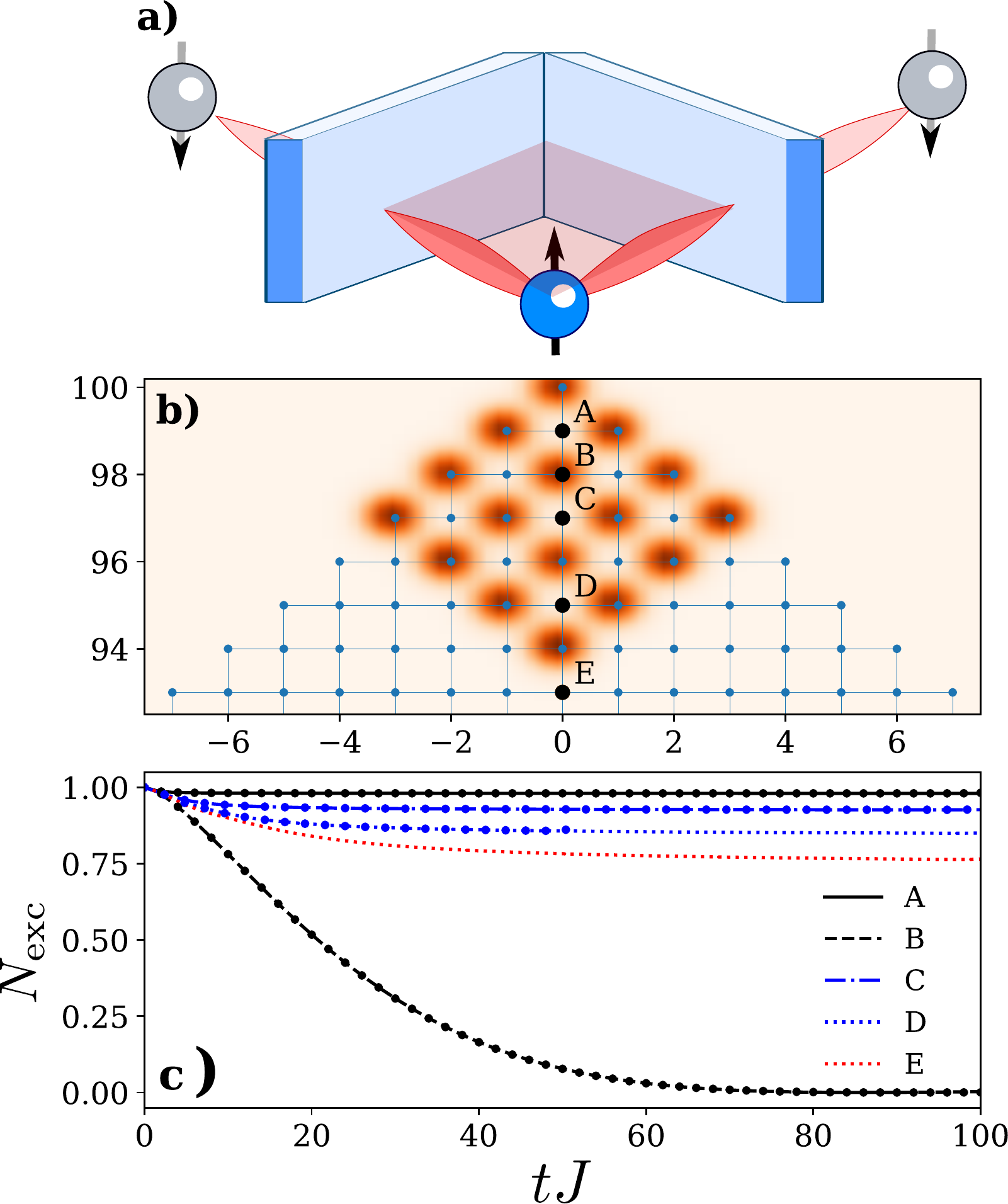}
    \caption{Formation of a corner state by spontaneous emission on a 2D rombic lattice with $30$ sites on each diagonal, for $g=0.01\Delta$ and $J=0.4\Delta.$ (a) Pictorical representation of the emitter and its afterimages. (b) Locations of the emitter in the corner of the photonic lattice (dots), coupling between photonic sites (lines) and distribution of photons (density plot), for a corner state generated by emitter $E.$ (c) Total excitation number $N_\text{excit}$ as a function of time, for different locations of the emitter, from $A$ to $E$.}
    \label{fig:2d-corner}
\end{figure}

\textit{Qubit-photon corner states in two-dimensions.--}
To obtain these phenomena in two dimensions, it is enough to consider the simpler generalization of the coupled cavity array to two dimensions, that is, disposing the resonators in a square geometry.  This model displays an energy dispersion given by:
\begin{equation}
\omega_\mathrm{2D}(\mathbf{k})=\omega_a+2J(\cos(k_x)+\cos(k_y))\,.
\end{equation}

At the middle of the band, $\omega(\mathbf{k})=\omega_a$, the isofrequencies are "nested", which means they are straight lines defined by $k_x\pm k_y=\pm,\mp \pi$. One of the consequences of such lines is that when an emitter is spectrally tuned to that energy, its emission becomes highly directional~\cite{gonzaleztudela17a,gonzaleztudela17b,galve17a}. This is what we will harness to induce the perfect trapping. As in the 1D case, the intuitive idea (see Fig.~\ref{fig:2d-corner}a) consists in placing the emitter in a position such that its directional emission is orthogonal to the bath boundaries, and their afterimages are out-of-phase with respect to the emission from the emitter.

Fig.~\ref{fig:2d-corner}b shows a proof-of-principle example of that mechanism. We have taken a square lattice and removed sites to form a reflective corner in a rhombus with $4\times 30^2$ sites. The quantum emitter is equidistant to its afterimages only when placed on the diagonal of the rhombus---positions $A$ to $E$ in the plot---. As in the 1D case, when we place the emitter on an odd site, such as B, it fails to acquire the right phase relation and decays releasing a photon into the lattice. However, for even positions (A, C, D, E) the emitter relaxes to a stationary state with high probability [cf. Fig.~\ref{fig:2d-corner}b]. In these states, the photon is trapped in a corner, avoiding the quantum emitter. Fig.~\ref{fig:2d-corner}b shows a density plot of a trapped photon that is anchored by a quantum emitter at position E. As in the 1D case, we have a very good agreement between DMRG and the single-photon polaron Hamiltonian for the rhombus. However, since the DMRG is working with a reduced number of modes (up to four per bath site) it allows the simulation of larger lattices---see Fig.\ \ref{fig:2d-corner}c, where the DMRG uses $400^2$sites---, and even moving to higher dimensional scenarios as we will show next.

\begin{figure}[!t]
	\centering
	\includegraphics[width=\linewidth]{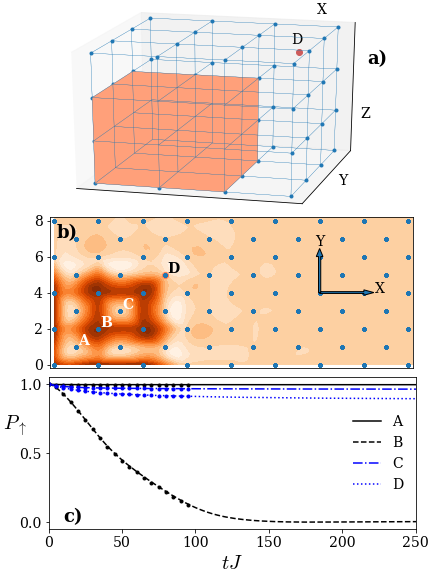}
	\caption{(a) A cube of light in a corner state trapped by a quantum emitter at position D $(x=y=z=5)$ on the diagonal of a BCC photonic lattice. (b) Density of photons on the corner state, as seen from above. (c) Probability of creating a corner state for emitters at A, B, C and D (respectively $x=y=z\in\{1,2,3,5\}$), for $J=0.4\Delta, g=0.005\Delta$ in Eq.~\eqref{eq:spin-boson}. Dots are simulations using the DMRG algorithm with 400 Lanczos states.}
	\label{fig:3d-corner}
\end{figure}

\textit{Qubit-photon corner states in 3D--}
In the three-dimensional case there are many different geometries in which the resonators can be disposed, but not all of them are suitable for our purposes. Using the intuition developed in Ref.~\cite{gonzaleztudela18d}, we choose a body-centered-cubic lattice in which each resonator is connected to four nearest neighbours. This model has an energy dispersion:
\begin{align}
\omega_{\mathrm{3D}}(\mathbf{k})=\omega_a+2J\Big[\cos(k_x)+\cos(k_y)+\\
+\cos(k_z)+\cos(k_x+k_y+k_z)\Big]\,,
\end{align}
with nested equifrequencies that yield highly collimated emission in 3 directions. This is especially well-suited to provide reflection in 3D corners. Other geometries, like the cubic-simple lattices, also display collimated emission but in more directions~\cite{gonzaleztudela18d}, such that they are not adequate for the desired goal.

In Fig.~\ref{fig:3d-corner} we provide a proof-of-principle numerical confirmation of the trapping for a lattice with \alex{$N=?$} and $g/\Delta=0.1$. Fig.~\ref{fig:3d-corner}a) shows the 3D photon distribution of a qubit-photon corner state when placed in the position denoted by the red dot (D), while in Fig.~\ref{fig:3d-corner}b) we plot an horizontal cut of this distribution. Finally, in Fig.~\ref{fig:3d-corner}c) we plot the probability of exciting the BIC as a function of time for the positions A-D depicted in Fig.~\ref{fig:3d-corner}b) comparing again the polaron Hamiltonian (lines) and chain-mapped DMRG (dots). Here again, we see the difference between the positions A, C, and D, where the phase relation with the afterimages is the right one, compared the B situation where the photon is not trapped, and BIC probability is very small.

\paragraph{Discussion.--} Summing up, in this work we have shown that a single quantum emitter can trap a photon in any dimension. The emitter must be placed in a photonic crystal-like medium, with the right separation from the reflective boundaries of the medium. Under such conditions, the emitter interferes destructively with the afterimages reflected by the boundaries, generating a bound-state-in-the-continuum that we denote as \textit{qubit-photon corner state.} We have shown evidence of this effect in 1D, 2D and 3D, from the rotating-wave approximation to the ultra-strong coupling regimes. 

As an outlook, let us mention that when several qubits are placed at the positions that form the qubit-photon corner states, they are effectively decoupled from environment, but still can interact coherently through the overlap of their photonic component. This opens new opportunities to design decoherence-free quantum gates~\cite{paulisch16a,kockum18a} in higher dimensions.

\begin{acknowledgments}
J.J.G.-R. and A.G.-T. acknowledge support from Project PGC2018-094792-B-I00~(MCIU/AEI/FEDER, UE), CSIC Research Platform PTI-001, and CAM/FEDER Project No.~S2018/TCS-4342~(QUITEMAD-CM). A.E.F. acknowledges the U.S. Department of Energy, Office of Basic Energy Sciences for support under grant No.~DE-SC0019275.
\end{acknowledgments}

%

\end{document}